\documentclass[fleqn,twoside]{article}
\usepackage[headings]{espcrc2}
\usepackage{xspace}
\usepackage{url}
\usepackage{relsize}
\usepackage{booktabs}
\usepackage{graphics}
\usepackage{pstricks,pst-node,pst-text,pst-3d}

\readRCS
$Id: espcrc2.tex,v 1.2 2004/02/24 11:22:11 spepping Exp $
\usepackage{graphicx}
\usepackage[figuresright]{rotating}

\usepackage{relsize}
\def\babar{\mbox{\slshape B\kern-0.1em{\smaller A}\kern-0.1em
    B\kern-0.1em{\smaller A\kern-0.2em R}}}

\newcommand{\AmS}{{\protect\the\textfont2
  A\kern-.1667em\lower.5ex\hbox{M}\kern-.125emS}}

\newcommand{\ev}{\ensuremath{\mathrm{\,e\kern -0.1em V}}\xspace}
\newcommand{\gev}{\ensuremath{\mathrm{\,Ge\kern -0.1em V}}\xspace}
\newcommand{\tev}{\ensuremath{\mathrm{\,Te\kern -0.1em V}}\xspace}

\def\nb         {\ensuremath{{\rm \,nb}}\xspace}

\def\invfb   {\ensuremath{\mbox{\,fb}^{-1}}\xspace}
\def\invab   {\ensuremath{\mbox{\,ab}^{-1}}\xspace}

\def\epem       {\ensuremath{e^+e^-}\xspace}
\def\bbbar {\ensuremath{b\overline b}\xspace}
\def\tautau     {\ensuremath{\tau^+\tau^-}\xspace}

\def\BR              {{\cal{B}}}
\def\BRul            {{\cal{B}}_{\mathrm{UL}}}
\def\BRulninety      {{\cal{B}}_{\mathrm{UL}}^{90}}

\def\electron   {\ensuremath{e}\xspace}

\def\piz   {\ensuremath{\pi^0}\xspace}

\newcommand{\etapr}{\ensuremath{\eta^{\prime}}\xspace}

\def\eff                       {\ensuremath{\varepsilon}\xspace}

\def\tenseven                  {\ensuremath{\times 10^{-7}}\xspace}
\def\teneight                  {\ensuremath{\times 10^{-8}}\xspace}

\def\mtau       {\ensuremath{\tau}\xspace}

\def\pzero                     {\ensuremath{P^0}\xspace}
\def\tautolpz                  {\ensuremath{\tau^\pm \to \ell^\pm \pzero}\xspace}

\def\tautolpiz                 {\ensuremath{\tau^\pm \to \ell^\pm \piz}\xspace}
\def\tautoepiz                 {\ensuremath{\tau^\pm \to \electron^\pm \piz}\xspace}
\def\tautompiz                 {\ensuremath{\tau^\pm \to \mu^\pm \piz}\xspace}

\def\tautoleta   {\ensuremath{\tau^\pm \to \ell^\pm \eta}\xspace}
\def\tautoeeta   {\ensuremath{\tau^\pm \to \electron^\pm \eta}\xspace}
\def\tautometa   {\ensuremath{\tau^\pm \to \mu^\pm \eta}\xspace}

\def\BRtaumeta   {\ensuremath{\BR(\tau^\pm \to \mu^\pm \eta)}\xspace}

\def\tautoletap   {\ensuremath{\tau^\pm \to \ell^\pm \etapr}\xspace}
\def\tautoeetap   {\ensuremath{\tau^\pm \to \electron^\pm \etapr}\xspace}
\def\tautometap   {\ensuremath{\tau^\pm \to \mu^\pm \etapr}\xspace}

\def\gaga  {\ensuremath{\gamma\gamma}\xspace}  %% changed from \gg, which is >>
\def\ptogg   {\ensuremath{\piz \to \gaga}\xspace}
\def\etogg   {\ensuremath{\eta \to \gaga}\xspace}
\def\etoppp  {\ensuremath{\eta \to \pi^+\pi^-\piz}\xspace}
\def\eptoppe {\ensuremath{\etapr \to \pi^+\pi^-\eta}\xspace}
\def\eptorg {\ensuremath{\etapr \to \rho^0\gamma}\xspace}

\newcommand{\taumg}      {\ensuremath{\tau^{\pm} \to \mu^{\pm} \gamma}\xspace}
\newcommand{\taueg}      {\ensuremath{\tau^{\pm} \to e^{\pm} \gamma}\xspace}
\newcommand{\taulg}      {\ensuremath{\tau^{\pm} \to \ell^{\pm} \gamma}\xspace}

\newcommand{\BRtaumg}    {\ensuremath{\BR(\taumg)}\xspace}
\newcommand{\BRtaueg}    {\ensuremath{\BR(\taueg)}\xspace}

\newcommand{\Nbkg}      {\ensuremath{N_{\rm bkg}}}
\newcommand{\Nobs}      {\ensuremath{N_{\rm obs}}}
\newcommand{\Nul}       {\ensuremath{N_{\rm UL}^{90}}}

\title{Searches for lepton flavor violating decays
$\tau^\pm\to\ell^\pm\gamma$, $\tau^\pm\to \ell^\pm P^0$
(where $\ell^- = e^-, \mu^-$, and $P^0$ = $\pi^0, \eta, \eta^{\prime}$)
at B-Factories: Status and Combinations}

\author{Swagato Banerjee\address{Department of Physics and Astronomy, University of Victoria,\\
                                  P.O. Box 3055, Victoria, British Columbia, V8W 3P6 Canada.}%
\thanks{To appear in the proceedings of 9$^{th}$ International Workshop on Tau-Lepton Physics (Tau06)
 held at Pisa, Italy, from 19$^{th}$ to 22$^{nd}$ September 2006.}
}
       
\runtitle{\taulg/\pzero}
\runtitle{Searches for lepton flavor violating decays
  $\tau^\pm\to\ell^\pm\gamma$, $\tau^\pm\to \ell^\pm P^0$}
\runauthor{Swagato Banerjee}

\begin{document}

\begin{abstract}
The present experimental status of searches for lepton flavor violating decays of the $\tau$ lepton 
to a lighter mass lepton and a photon or a pseudoscalar meson is presented. 
The results obtained are from $e^+e^-$ annihilation data collected 
at a center-of-mass energy near 10.58 GeV by the \babar\ and the Belle detectors.
In order to benefit from larger available datasets, the first $\babar-$Belle combinations of upper limits 
on the branching fractions of lepton flavor violating $\tau$-decays are obtained.
\vspace{1pc}
\end{abstract}

\maketitle

\section{Introduction}
The recent discovery of neutrino mixing~\cite{NuOsc} implies that lepton flavor violation (LFV) occurs.
LFV in charged decays have not yet been observed,
although they have long been identified as an unambiguous signature of new physics.

The Standard Model (SM) extended to include finite mass differences from neutrino mixing 
imply $\tau - \mu$ mixing is generated at the one-loop level and 
thus suppressed by a factor of $(m^2_\nu/m^2_W)^2$,
%\BRtaumg \sim {\cal{O}} (10^{-54})$~\cite{Aubert:2005ye},
which is many orders of magnitude below the experimental sensitivity.
However, many new theories, for example 
SM with additional heavy right-handed Majorana neutrinos or 
with left-handed and right-handed neutral isosinglets~\cite{Cvetic:2002jy};
mSUGRA models with right handed neutrinos introduced 
via the seesaw mechanism~\cite{Ellis:1999uq,Ellis:2002fe};
supersymmetric models with Higgs exchange~\cite{Dedes:2002rh,Brignole:2003iv} 
or SO(10) symmetry~\cite{Masiero:2002jn,Fukuyama:2003hn};
or technicolor models with non-universal Z$^\prime$ exchange~\cite{Yue:2002ja},
allow for \taulg decays, where $\ell^- = e^-, \mu^-$, at the level of $\sim {\cal{O}} (10^{-10}-10^{-7})$.

Large neutrino mixing~\cite{NuOsc} could potentially 
induce large mixing between the supersymmetric partners of the leptons.
While some scenario's predict higher rates for \taumg decays, others, 
for example with inverted hierarchy rather than normal hierarchy 
of masses of supersymmetric partners of leptons,
predict~\cite{Ellis:2002fe} higher rates for \taueg decays.

Semi-leptonic neutrinoless decays involving pseudoscalar mesons like \tautolpz, 
where $\pzero = \piz, \eta, \etapr$, 
are likely candidates for LFV in supersymmetric models~\cite{SusyLFV},
for example, arising out of exchange of CP-odd pseudoscalar neutral Higgs boson,
which are further enhanced by color factors associated with these decays.
The large coupling of Higgs at $s\bar{s}$ vertex enhances final state containing the $\eta$ meson,
giving a prediction of  $\BRtaumeta : \BRtaumg = 8.4 : 1.5$~\cite{Sher:2002ew}.
Some models with heavy Dirac neutrinos~\cite{Gonzalez-Garcia:1991be,Ilakovac:1999md},
two Higgs doublet models, R-parity violating supersymmetric models,
and flavor changing $Z^\prime$ models with non-universal couplings~\cite{Li:2005rr}
allow for observable parameter space of new physics~\cite{Black:2002wh},
while respecting the existing experimental bounds at the level of $\sim {\cal{O}}(10^{-7})$.

Present generation $\epem$ B-factories operating around the $\Upsilon$(4S) 
resonance also serve as $\tau$-factories,
because the production cross-sections of $\sigma_{\bbbar} : \sigma_{\tautau} = 1.1 : 0.9 \nb$ are comparable to each other 
at center-of-mass (CM) energy near 10.58 \gev.
Recent results from the searches of \taulg/\pzero are presented here,
along with a first attempt at combining the upper limits from the \babar\ and the Belle experiments.

\section{B-Factories: Status}

In \epem annihilations, \taulg/\pzero decays have 2 characteristic features:
the energy of tau-daughters is close to the half the CM energy,
and the total invariant mass of the daughters is close to the mass of \mtau-lepton.
Sometimes, the mass of the daughters are calculated with total energy constrained to be half the CM energy,
in order to improve upon the resolution of the mass variable.

The general strategy for searches of such decays is 
to define a signal region in the energy-mass plane of the \mtau-daughters
(typically $2\sigma$ region), and optimize upon a set of selection criteria 
to reduce the contribution to background events from well-known SM processes.
The principal sources of background are radiative di-muon or Bhabha events,
along with some irreducible contribution from \mtau-pair events having hard ISR
with $\tau^\pm\to\ell^\pm\nu\bar{\nu}$ or $\tau^\pm\to\pi^\pm\piz\bar{\nu}$ decays for \taulg\ or \tautolpz\ searches, respectively.

After applying the selection criteria, the signal efficiency (\eff) and 
expected background events (\Nbkg) are estimated inside the signal region.
Finally, the number of observed events (\Nobs) are counted (or fitted) from the data.
As yet, no signal has been observed in any of the searches.
So, an upper limit at 90\% confidence level (C.L.) is obtained as:
$\BRulninety = \Nul/(N_{\tau} \eff)$, where 
\Nul is the 90\% confidence level (C.L.) upper limit
on the number of signal events inside the signal box, and
$N_{\tau} = 2 {\cal{L}} \sigma_{\tautau}$
is the number of \mtau-decays studied from a sample of data with a luminosity of ${\cal{L}}$.

For the \tautolpz searches, the pseudoscalar mesons are reconstructed in the following decay modes:
\ptogg for \tautolpiz,
\etogg and \etoppp (\ptogg) for \tautoleta,
\eptoppe (\etogg) and \eptorg for \tautoletap.
In these searches, the signal efficiency (\eff) in the above formula for $\BRulninety$ is the product of
the branching fraction ($\BR$) and reconstruction efficiency ($\xi$) of the $P^0$ decay mode under consideration.
To obtain a combined upper limit with $\eta$ and $\etapr$ decays,
the observed and expected background events
and the signal efficiencies are added using $\BR\xi = (\BR_1 \times \xi_1 + \BR_2 \times \xi_2)$,
where $\BR_1$, $\BR_2$ are the respective branching fractions
and $\xi_{1}$, $\xi_{2}$ are the corresponding reconstruction efficiencies.

Searches for LFV in $\taulg$ and $\tautolpz$ decays have been performed 
with ${\cal{L}}$ = 232.2\invfb and 339\invfb by the \babar\ experiment~\cite{Aubert:2005ye,Aubert:2005wa,Aubert:2006cz}, and
with ${\cal{L}}$ = 535\invfb   and 401\invfb by the Belle experiment~\cite{Abe:2006sf,Abe:2006qv}, respectively.
The results for these searches are summarized in Table 1, where
$\BR(\ptogg) = (98.80\pm0.03)\%$, 
$\BR(\etogg) = (39.38\pm0.26)\%$, 
$\BR(\etoppp,\ptogg)  = (22.43\pm0.40)\%$,
$\BR(\eptoppe,\etogg) = (17.52\pm0.56)\%$, and 
$\BR(\eptorg) = (29.40\pm0.90)\%$ 
have been used~\cite{Yao:2006px} for combining modes.

\section{B-Factories: Combinations}
The results from the \babar\ and the Belle experiments are 
combined by adding the number of observed and expected background events along with their symmetrized errors, 
respectively, inside the signal region. 
%Asymmetric errors are averaged, except when the contribution becomes negative.
%in which case the lower errors are set to zero.
The signal efficiencies are averaged using weights than account for the
integrated luminosity for the corresponding searches.
This is the most conservative approach given that the signal and background PDFs are different for 
similar searches by different experiments, because of differing resolutions on reconstructed mass and energy among the experiments.
Thus, in this procedure, the distribution of both the signal and the background events inside the signal
region are taken to be uniform, and only the rates of signal and backgrounds events contribute to the combination.

Combined frequentist limits are calculated including 
all uncertainties using the technique of Cousins and Highland~\cite{Cousins:1992qz}
following the implementation of Barlow~\cite{Barlow:2002bk}.
In this technique, 10$^6$ MC samples are generated 
according to a Poisson distribution with mean $(s + b)$,
where the background $(b)$ and signal $(s)$ are each drawn randomly 
from Gaussian distributions describing their respective PDFs.
For the signal Gaussian, the mean and the standard deviation are $2{\cal{L}}\sigma_{\tau\tau}\BRul\eff$
and its uncertainty, with the latter determined by propagation of errors.
The branching ratio $(\BRul)$ is varied until we find a value for which 10\% of the samples yield a number of events
less than that observed in the combined data. 
At 90\% C.L. this procedure gives the combined upper limit observed for each decay mode.
\clearpage

\begin{table*}[!h]
\caption{
The luminosities, the efficiencies,
the number of expected and observed background events inside the signal region,
and the expected and observed upper limits on the branching fractions at 90\% C.L.($\BRulninety$)
for the searches from the \babar\ ~\cite{Aubert:2005ye,Aubert:2005wa,Aubert:2006cz}
and the Belle~\cite{Abe:2006sf,Abe:2006qv} experiments and their combinations as described in the text.
An empty cell indicates that the respective value has not been published.}

\label{table:1}
\newcommand{\m}{\hphantom{$-$}}
\newcommand{\cc}[1]{\multicolumn{1}{c}{#1}}
\renewcommand{\tabcolsep}{.7pc} % enlarge column spacing
\renewcommand{\arraystretch}{1.2} % enlarge line spacing
\begin{center}\smaller
\begin{tabular}{lcccccc}\toprule
                 & Luminosity & Efficiency & \multicolumn{2}{c}{Background events} & \multicolumn{2}{c}{$\BRulninety$ (\teneight)} \\

                 &(\invfb)&     {(\%)}     & {Expected}             & {Observed}     & {Expected} & {Observed} \\
\hline

\multicolumn{7}{c}{\taueg}\\\hline
\babar\ & 232.2 & {$4.70\pm0.29$} & {$1.9\pm0.4$}          & {1}            & 12         & 11   \\
Belle & 535.0 & {$2.99\pm0.13$} & {$5.14^{+2.6}_{-1.9}$} & {5}            &            & 12   \\
\babar\ \&                                                                                
Belle & 767.2 & {$3.51\pm0.13$} & {$7.0\pm2.3$}          & {6}            & 12         &  9.4 \\
\hline
                                                                                                  
\multicolumn{7}{c}{\taumg}\\\hline                                                              
\babar\ & 232.2 & {$7.42\pm0.65$} & {$6.2\pm0.5$}          & {4}            & 12         &  6.8 \\
Belle & 535.0 & {$5.07\pm0.20$} & {$13.9^{+3.3}_{-2.6}$} & {10}           &            &  4.5 \\
\babar\ \&                                                                                
Belle  & 767.2 & {$5.78\pm0.24$} & {$20.1\pm3.0$}         & {14}           & 11         &  1.6 \\
\hline
                                                                                                  
\multicolumn{7}{c}{\tautoepiz}\\\hline                                                          
\babar\ & 339.0 & {$2.83\pm0.25$} & {$0.17\pm0.04$}        & {0}            & 14         & 13   \\
Belle & 401.0 & {$3.93\pm0.18$} & {$0.20\pm0.20$}        & {0}            &            &  8.0 \\
\babar\ \&                                                                                
Belle  & 740.0 & {$3.42\pm0.15$} & {$0.37\pm0.20$}        & {0}            & 5.7        &  4.4 \\
\hline
                                                                                                  
\multicolumn{7}{c}{\tautompiz}\\\hline                                                          
\babar\ & 339.0 & {$4.75\pm0.37$} & {$1.33\pm0.15$}        & {1}            & 11         & 11   \\
Belle & 401.0 & {$4.53\pm0.20$} & {$0.58\pm0.34$}        & {1}            &            & 12   \\
\babar\ \&                                                                                
Belle & 740.0 & {$4.63\pm0.20$} & {$1.91\pm0.37$}        & {2}            &  5.4       &  5.8 \\
\hline
                                                                                                  
\multicolumn{7}{c}{\tautoeeta}\\\hline                                                          
\babar\ & 339.0 & {$2.12\pm0.20$} & {$0.22\pm0.05$}        & {0}            & 19         & 16 \\
Belle & 401.0 & {$2.86\pm0.14$} & {$0.78\pm0.59$}        & {0}            &            & 9.2 \\
\babar\ \&                                                                                
Belle  & 740.0 & {$2.52\pm0.12$} & {$1.00\pm0.59$}        & {0}            &  8.9       & 4.5 \\
\hline
                                                                                                  
\multicolumn{7}{c}{\tautometa}\\\hline                                                          
\babar\ & 339.0 & {$3.59\pm0.41$} & {$0.75\pm0.08$}        & {1}            & 13         & 15  \\
Belle & 401.0 & {$4.06\pm0.20$} & {$0.64\pm0.38$}        & {0}            &            &  6.5 \\
\babar\ \&                                                                                
Belle & 740.0 & {$3.85\pm0.22$} & {$1.39\pm0.38$}        & {1}            &  6.0       &  5.1 \\
\hline
                                                                                                  
\multicolumn{7}{c}{\tautoeetap}\\\hline                                                         
\babar\ & 339.0 & {$1.53\pm0.16$} & {$0.12\pm0.03$}        & {0}            & 26         & 24   \\
Belle & 401.0 & {$2.15\pm0.12$} & {$0.00^{+0.41}_{-0.00}$}&{0}            &            & 16   \\
\babar\ \&                                                                                
Belle & 740.0 & {$1.86\pm0.10$} & {$0.12^{+0.41}_{-0.03}$}&{0}            &  9.8       &  9.0 \\
\hline
                                                                                                  
\multicolumn{7}{c}{\tautometap}\\\hline                                                         
\babar\ & 339.0 & {$2.18\pm0.26$} & {$0.49\pm0.04$}        & {0}            & 20         & 14   \\
Belle & 401.0 & {$2.45\pm0.13$} & {$0.23^{+0.33}_{-0.23}$}&{0}            &            & 13   \\
\babar\ \&                                                                                
Belle  & 740.0 & {$2.33\pm0.14$} & {$0.72\pm0.28$}        & {0}            &  8.9       &  5.3 \\\bottomrule
\end{tabular}
\end{center}
\end{table*}
\clearpage

\noindent In order to calculate the expected upper limit in absence of a signal,
the upper limit for each guess of the observed events is weighted by a Poisson
distribution centered around the number of expected background events and averaged.

The results of these combinations for the expected and the observed upper limits 
at 90\% C.L. $(\BRulninety)$ are summarized in Table 1.
In the case of \taumg, the observed limit is significantly lower than the
expected limit, because both experiments observe less events than expected~\cite{poisson}.

\section{B-Factories: Projections}

Depending upon the nature of backgrounds contributing to a given search, 
two scenarios can be envisioned while extrapolating to higher luminosities.
In the first scenario, there is no significant background contribution,
and the selection procedure may be optimized to give ${\cal{O}}(1)$ expected background events,
and so, $\BRulninety \propto 1/{\cal{L}}$.
In the other scenario, there is an irreducible background contribution, so that
$\Nul \propto \sqrt{\Nobs}$ grows as $\sqrt{\cal{L}}$, and so, $\BRulninety \propto 1/\sqrt{\cal{L}}$.

Taking as baseline the expected upper limit of $\BRtaumg = 1.2 \tenseven$ 
from search by the \babar\ experiment with ${\cal{L}} = 232.2 \invfb$,
one can expect upper limits of $\BRtaumg = 1.4 \teneight$ or $4.1 \teneight$
in scenarios with background free or background dominated search, respectively,
in a couple of years when both the \babar\ and the Belle experiments have accumulated 
total data with luminosity of ${\cal{L}} = 2 \invab$. If no signal is found,
the observed limit may be worse or better than the expected limit, 
depending upon if the observed number of data events is more or less than the
expected number of background events, as evident from the trend in Table 1.

\section{Constraints on new physics}
Improved limits on LFV in \mtau decays give several interesting constraints on new physics.
Two such constraints are discussed here: one from \BRtaumg and one from \BRtaumeta.

In supersymmetric models with SU(5) symmetry for Grand Unification Theories,
flavor changing right-handed currents induce correlations between 
possibly enhanced rates for \taumg decays and CP-asymmetry in b-s penguins~\cite{Hisano:2003bd}.
The HFAG summer 2006 average~\cite{hfag:phik} of \babar\ ~\cite{Aubert:2006av} 
and Belle~\cite{Chen:2006nk} measurements of $S(B \to \phi K^0) = (0.39 \pm 0.18)$ 
is already very promising.
However, an observed upper limit of \BRtaumg $<$ 1.6 \teneight from this combination 
of results from the \babar\ and the Belle experiments 
with a combined ${\cal{L}} = 767.2 \invfb$ is more constraining, as shown in Figure 1.
The contours are shown for fixed values of universal gaugino mass $m_{\tilde{g}}$ = 400, 600, 800, and 1000 \gev,
with universal scalar mass $m_0 \in [200, 1000] \gev$,
the ratio of the vacuum expectation values of the two Higgs doublets $\tan\beta = 10$, 
the trilinear coupling $A_0 = 0$,
the mass of the right-handed neutrino $m_{\nu_R} = 5 \times 10^{14} \gev$, and
the mass of left-handed $\tau$-neutrino $m_{\nu_\tau} = 5 \times 10^{-2} \ev$
(for description of other parameters see Figure 2 in Reference~\cite{Hisano:2003bd}).

Mixing between left-handed smuons and staus allows one to translate the upper limit on \BRtaumeta 
to an exclusion plot in the $\tan\beta$ vs.\ $m_A$ plane,
where $m_A$ is the mass of the CP-odd pseudoscalar neutral Higgs boson.
The excluded regions at 95\% C.L. from this combination of results from
the \babar\ and the Belle experiments for \tautometa search ($<6.8\teneight$)
with right-handed neutrino mass = 10$^{14}$\gev introduced via the seesaw mechanism
in supersymmetric models~\cite{Sher:2002ew} are shown in Figure 1.
Also are shown the regions excluded by the \babar\ ~\cite{Aubert:2006cz}
and the Belle~\cite{Abe:2006qv} experiments individually.
These results are competitive with those obtained from the direct searches for 
Higgs $\to$ $b\bar{b}$, $\tautau$ decays by the CDF~\cite{cdf} and the D0~\cite{dzero} experiments, 
and complementary to the region excluded by the LEP experiments with a top quark mass of 174.3 \gev\cite{lephiggs},
for two common scenarios of stop-mixing benchmark models~\cite{Carena:1999xa}:
$m_h^{\rm{max}}$ and no-mixing obtained with the Higgs mass parameter $\mu = -200 \gev$
shown by darker and lighter shaded regions respectively.

\section{Summary}
An improvement of five order of magnitude in the upper limits for \BRtaueg and \BRtaumg
over the last twenty-five years~\cite{Yao:2006px} 
is shown in Figures 3 and 4, respectively.
The next five years promises to be the most interesting phase in this evolution,
when experiments approach closer to the predictions of different theoretical models.

\begin{figure}[b]
\begin{center}
\includegraphics[width=.98\columnwidth,height=.285\textheight]{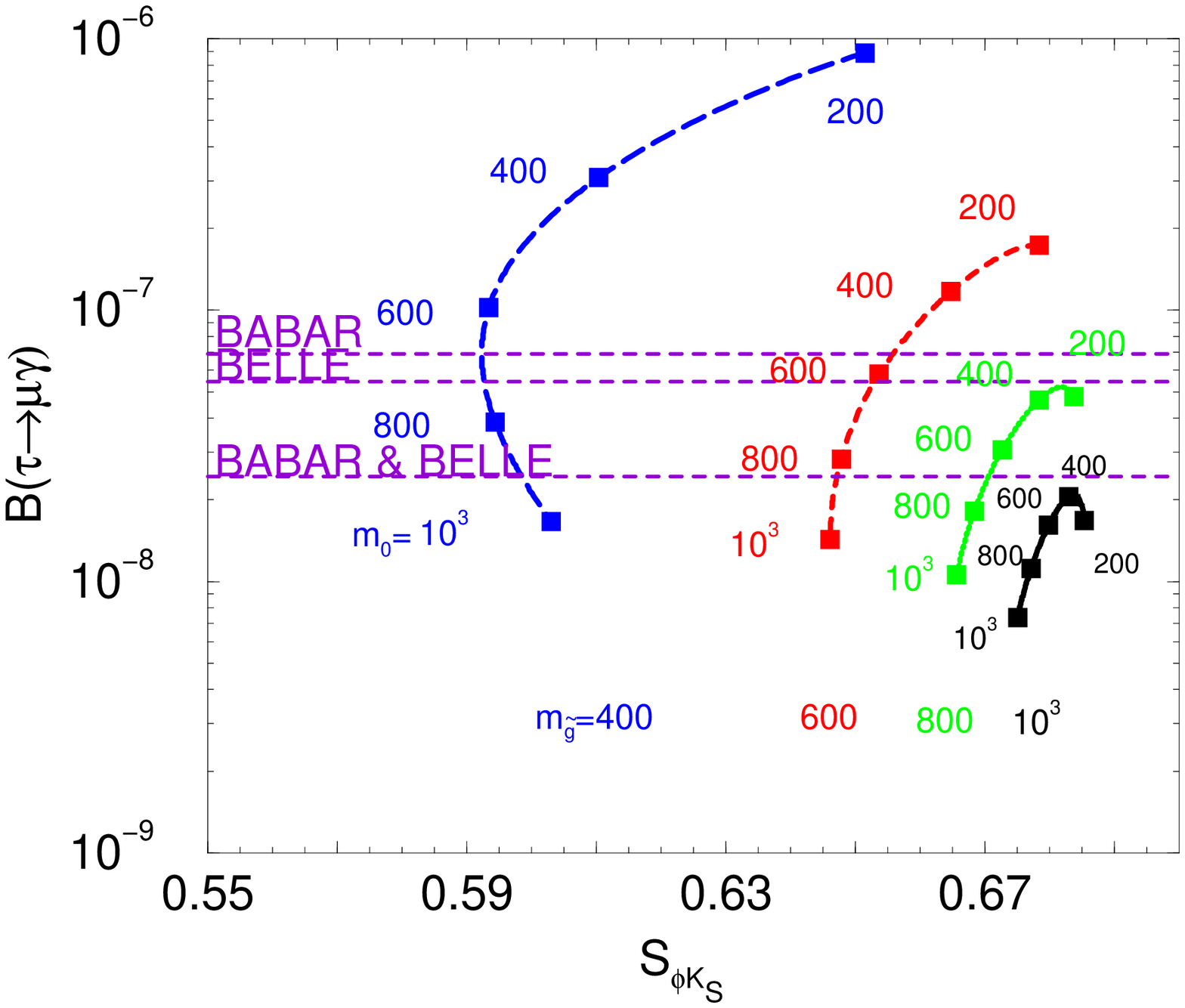}
\caption{\BRtaumg vs.\ $S(B \to \phi K^0)$ contours as function of $m_{\tilde{g}}$, $m_0$ (see text).}
\end{center}
\end{figure}

\begin{figure}[b]
\begin{center}
\includegraphics[width=.98\columnwidth,height=.285\textheight]{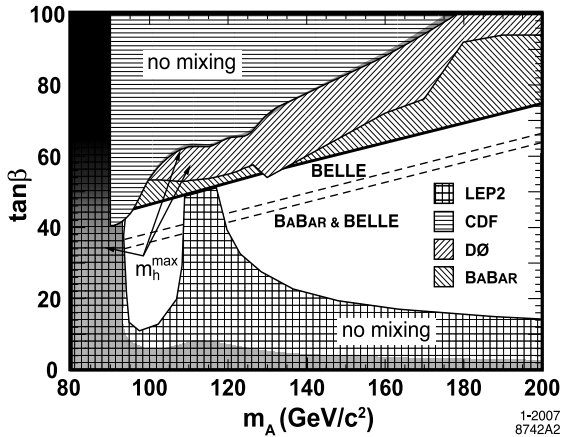}
\caption{Excluded regions in $\tan\beta$ vs.\ $m_A$ plane from search for \tautometa decays (see text).}
\end{center}
\end{figure}

\begin{figure}[t]
\begin{center}
\includegraphics[width=.98\columnwidth,height=.285\textheight]{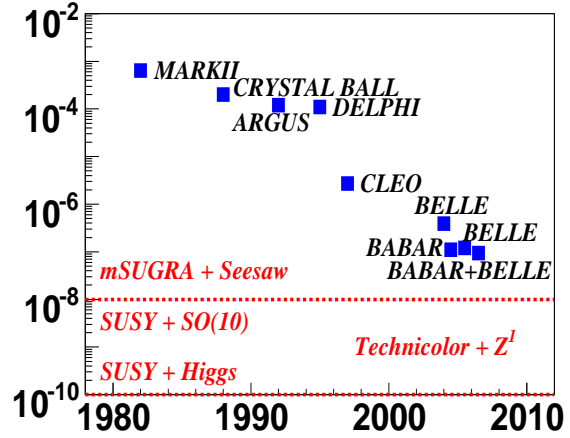}
\caption{Evolution of experimental bounds of \BRtaueg and some theoretical predictions.}
\end{center}
\end{figure}

%\vspace*{.5cm}
\begin{figure}[t]
\begin{center}
\includegraphics[width=.98\columnwidth,height=.285\textheight]{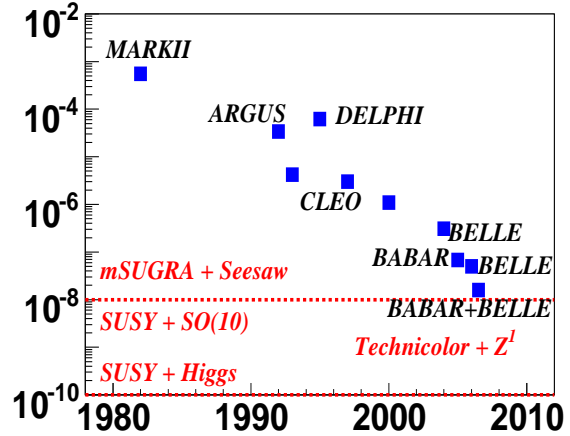}
\caption{Evolution of experimental bounds of \BRtaumg and some theoretical predictions.}
\end{center}
\end{figure}

\section*{Acknowledgments}

The author thanks the organizers of Tau06 
for the opportunity to present a first combination of upper limits 
on LFV in \mtau decays from the \babar\ and the Belle experiments. 
Discussions with
Kiyoshi Hayasaka,
Hisaki  Hayashii, 
Alberto Lusiani,
Takayoshi Ohshima,
Michael Roney and
Marc Sher
are gratefully acknowledged.

%%%%%%%%%%%%%%%%%%%%%%%%%%%%%%%%%%%%%%%%%%%%%%%%%%%%%%%%%%%%%%%%%%%%%%%%%%%%%%%%%%%%%%%%%%%%%%%


\begin{thebibliography}{9}
\bibitem{poisson}
The Poisson probability to observe less than or equal to 14 events
when 20.1 events are expected is 10\%.
\end{thebibliography}
\end{document}